\documentclass[12pt]{iopart}

\newcommand{\Tc}{T$_c$}
\newcommand{\CFS}{Cs$_x$Fe$_{2-y}$Se$_2$}

\usepackage{graphicx}
\usepackage{cite} 
\usepackage{multirow} 
\begin{document}

\title{Microstructural analysis of phase separation in iron chalcogenide superconductors}

\author{S C Speller$^1$, T B Britton$^1$, G M Hughes$^1$, A Krzton-Maziopa$^2$\footnote{On leave from Faculty of Chemistry, Warsaw University of Technology, 00-664, Warsaw, Poland}, E Pomjakushina$^2$, K. Conder$^2$, A T Boothroyd$^3$ and C R M Grovenor$^1$}

\address{$^1$ Department of Materials, University of Oxford, Oxford, OX1 3PH, UK}
\address{$^2$ Paul Scherrer Institute, CH-5232 Villigen PSI, Switzerland}
\address{$^3$ Department of Physics, University of Oxford, Oxford, OX1 3PU, UK}
\ead{susannah.speller@materials.ox.ac.uk}

\begin{abstract}
The  interplay between superconductivity, magnetism and crystal structure in iron-based superconductors is a topic of great interest amongst the condensed matter physics community as it is thought to be the key to understanding the mechanisms responsible for high temperature superconductivity.  Alkali metal doped iron chalcogenide superconductors exhibit several unique characteristics which are not found in other iron-based superconducting materials such as antiferromagnetic ordering at room temperature, the presence of ordered iron vacancies and high resistivity normal state properties.  Detailed microstructural analysis is essential in order to understand the origin of these unusual properties.  Here we have used a range of complementary scanning electron microscope based techniques, including high-resolution electron backscatter diffraction mapping, to assess local variations in composition and lattice parameter with high precision and sub-micron spatial resolution.  Phase separation is observed in the \CFS\ crystals, with the minor phase  distributed in a plate-like morphology throughout the crystal.  Our results are consistent with superconductivity occurring only in the minority phase.
\end{abstract}


\maketitle

\section{Introduction}
The recent discovery of superconductivity in a large family of iron-based pnictide and chalcogenide compounds provides great opportunities to improve the understanding of the mechanisms of high temperature superconductivity.  The crystal structures of the Fe-based compounds are similar to the cuprate superconductors, consisting of Fe-pnictide or Fe-chalcogenide layers responsible for superconductivity, usually separated by layers of other atoms.  Binary FeSe, with a transition temperature of 8K, is the simplest of the superconducting phases, consisting of stacked FeSe layers with no spacing atoms \cite{Fang:2008}. Its transition temperature can be increased to 14K by  substitution of about half of the Se atoms with larger Te atoms \cite{Yeh:2008}.  Increasing the fraction of Te further results in the development of antiferromagnetic (AFM) order coexisting with superconductivity over a narrow composition range.  It is not certain whether these different properties originate from spatially separated regions of the crystal, or whether homogeneous regions of the crystal exhibit both properties simultaneously \cite{Bendele:2010}, although recent microstructural studies suggest that a two-phase description is appropriate \cite{Speller:2011, Hu:2011}.  

In 2010 it was discovered that the introduction of potassium atoms between the FeSe layers in the crystal structure, producing a ternary compound with nominal composition of K$_x$Fe$_2$Se$_2$, significantly increases the superconducting transition temperature to $\approx$ 30K  \cite{Guo:2010}.  Subsequently a range of compounds in this family (A$_{x}$Fe$_{2-y}$Se$_2$ where A=K, Cs, Rb, Tl etc.) have been found to superconduct. The composition of these compounds are well-known to deviate from the ideal stoichiometry \cite{Mou:2011}, with Fe-vacancies introduced  into the structure owing to the restrictions on the valency of the iron atom.  At least five different types of iron ordering have been found in A$_{x}$Fe$_{2-y}$Se$_2$ compounds using both bulk techniques such as X-ray  and neutron diffraction and high resolution techniques such as TEM and STM microscopy, as discussed in a recent review article by Mou \etal \cite{Mou:2011}.  Experimentally determined phase diagrams  for Rb$_x$Fe$_{2-y}$Se$_2$ \cite{Tsurkan:2011} and K$_x$Fe$_{2-y}$Se$_2$ \cite{Yan:2012} indicate that AFM ordering occurs at temperatures above 500K over the composition range presented, with superconductivity co-existing with this AFM phase over a  narrow range of compositions.  Compounds with compositions either side of this region are insulating (or semiconducting), exhibiting different forms of AFM ordering and vacancy ordering schemes.  There is increasing evidence that the parent compound is the $\sqrt5$x$\sqrt5$ ordered Fe vacancy phase with composition A$_{0.8}$Fe$_{1.6}$Se$_{2}$, exhibiting AFM ordering and insulating properties \cite{Fang:2011, Yan:2012}.  In order to obtain the superconducting phase, extra Fe must be added, as Fe vacancies are considered to be detrimental to superconductivity \cite{Li:2011, Texier:2012}. 

SEM \cite{Ryan:2011}, TEM \cite{Li:2011, Wang:2011}, STM \cite{Li:2012} and nanofocused XRD studies \cite{Ricci:2011b} have all shown that phase separation exists on the nano-scale in crystals exhibiting large shielding fractions in magnetisation measurements, and this two phase nature is supported by muon-spin \cite{Shermadini:2012}, Mossbauer\cite{Ryan:2011} and, very recently, NMR results\cite{Texier:2012}.  Hu \etal have suggested that the apparently contradictory properties of these compounds, such as high \Tc\ values and large shielding fractions coupled with high electrical  resistivity and relatively small jump in specific heat, can be resolved if the crystals consist of a sub-micron scale ``aerogel'' network of minority conducting phase (superconducting below $\approx$ 30K) within an insulating, antiferromagnetic matrix \cite{Hu:2012}.  


This paper investigates in detail the microstructure of a series of three \CFS\  crystals with different superconducting properties in an attempt to gain further understanding of the factors affecting their properties.  Their microstructures are compared to those of Fe(Se,Te) single crystals in which superconductivity and magnetism are found to co-exist.  In particular,  high-resolution electron backscatter diffraction (HR-EBSD) has been used used to map  variations in lattice parameter with excellent precision and sub-micron spatial resolution.

\section{Experimental Methods}
The \CFS\ single crystals used in this study have been grown by the Bridgman process detailed previously \cite{Krzton-Maziopa:2011}.  Samples are prepared very carefully for microscopy because they are highly air-sensitive, with the highly mobile Cs ions quickly diffusing to the surface and reacting with oxygen, resulting in microstructures and local chemical compositions which are not characteristic of the bulk.   The results presented here are all carried out on samples freshly cleaved on the (001) plane immediately prior to insertion into the vacuum chamber of the  microscope, minimising exposure to air to under 10 seconds.  
Energy dispersive X-ray (EDX) analysis has been carried out in a JEOL 6300 scanning electron microscope (SEM) using Oxford Instruments INCA software with internal standards for quantisation.  Extensive EDX analysis has been carried out on at least two different flakes taken from each crystal.  
Cross-sections have been produced using a Zeiss NVision focussed ion beam microscope, with low-loss backscattered electron imaging carried out in-situ using the Gemini column on this instrument.
High-resolution electron backscatter diffraction (HR-EBSD) analysis has been carried out in a JEOL 6500F field emission gun SEM, with off-line pattern analysis carried out using CrossCourt v3 (BLG productions) software.  In this technique (developed by Wilkinson et al. \cite{Wilkinson:2006}), image correlation is used to determine local elastic (i.e. lattice) strain variations and lattice rotations to a precision of 1x10$^{-4}$ in strain and 0.006$^{\circ}$ in rotation from the displacement gradient tensor.  However, it is only possible to directly measure the variations in differences in normal strain components (e.g. $\varepsilon_{33}-\varepsilon_{11}$) because hydrostatic dilation does not affect the interplanar angles measured using this image correlation technique.   In this work variations in the anisotropy of the crystals (c/a ratio) is presented, as this ratio can be measured directly with the HR-EBSD technique without knowledge of the elastic properties of the material (as detailed in \cite{Speller:2011}).  The absolute c/a values are calculated by assuming the modal c/a value for the map is equal to the bulk c/a value measured by full-pattern refinement of powder X-ray diffraction (XRD).  Superconducting and magnetic property measurements of these crystals have been reported previously \cite{Krzton-Maziopa:2012,Shermadini:2011,Pomjakushin:2011}.  
 
\section{Microstructural analysis of \CFS\ crystals}
In this section we discuss detailed microstructural analysis carried out on three different  \CFS\ single crystal samples; samples K100 and K47 exhibit both superconductivity and antiferromagnetic (AFM) ordering, whilst sample K73 is insulating with AFM ordering as summarised in table \ref{table:samples}.  

\begin{table}[h!]
 \begin{center}
 \caption{Details of the bulk properties of the \CFS\ crystals used in this study.}
   \begin{tabular}{| l | l | l | l | l |}
   \hline
   \multirow{2}{*}{Sample} &
   \multicolumn{2}{c|}{Composition} &
   \multirow{2}{*}{c/a ratio} &
   \multirow{2}{*}{\Tc} \\
   \cline{2-3}
   & nominal & micro XRF & &\\
   \hline
  K100 & Cs$_{0.8}$Fe$_{1.9}$Se$_{2}$ &  Cs$_{0.72}$Fe$_{1.57}$Se$_{2}$ & 3.8621(10) & 27K \\
K47 & Cs$_{0.87}$Fe$_{2.04}$Se$_{2}$ &  Cs$_{0.73}$Fe$_{1.57}$Se$_{2}$ & 3.8565(9)& 23K\\
K73 & Cs$_{0.8}$Fe$_{1.8}$Se$_{2}$ &  Cs$_{0.68}$Fe$_{1.52}$Se$_{2}$ &3.8677(8) & - \\
   \hline
   \end{tabular}
 \end{center}

\label{table:samples}
\end{table}

\subsection{Morphology}
\begin{figure}[h!]
  \begin{center}
   \includegraphics[width=1\textwidth]{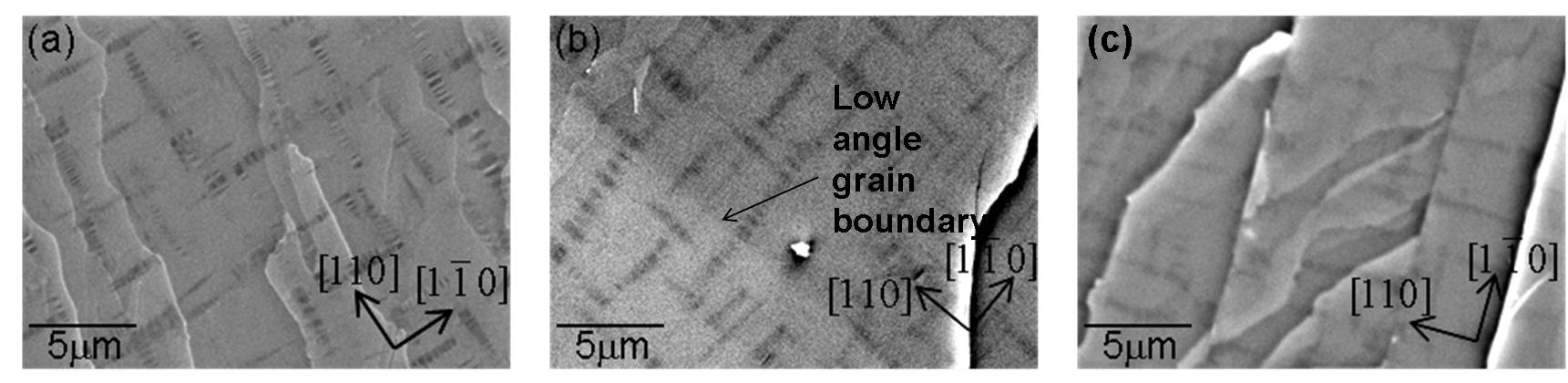}

  \end{center}
  \caption{Scanning electron micrographs showing the two phase nature of \CFS\ crystals (a) K100, (b) K47 and (c) K73.}
\label{fig:SE}
\end{figure}

\begin{figure}[h!]
  \begin{center}
   \includegraphics[width=0.45\textwidth]{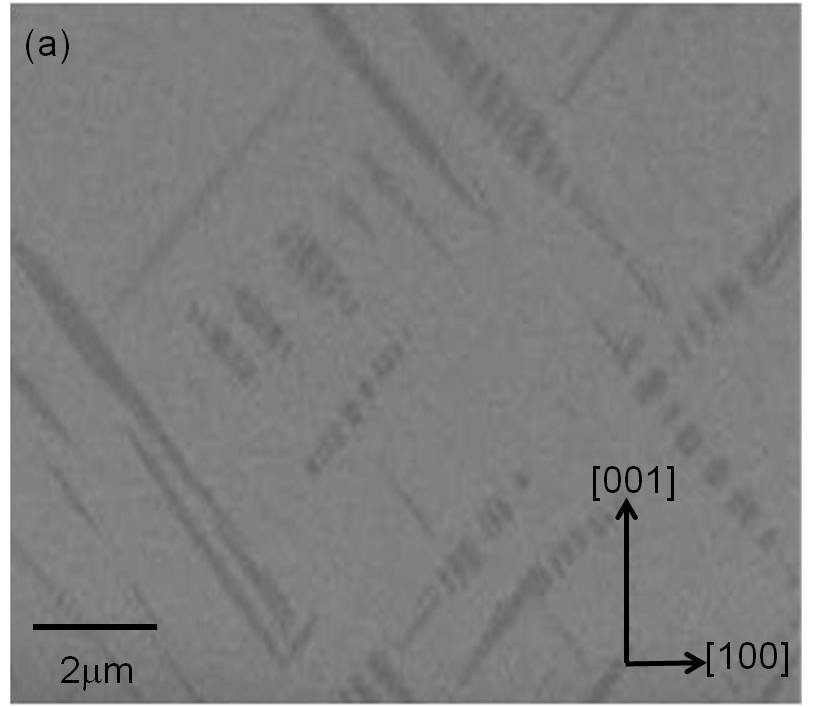}
 \includegraphics[width=0.45\textwidth]{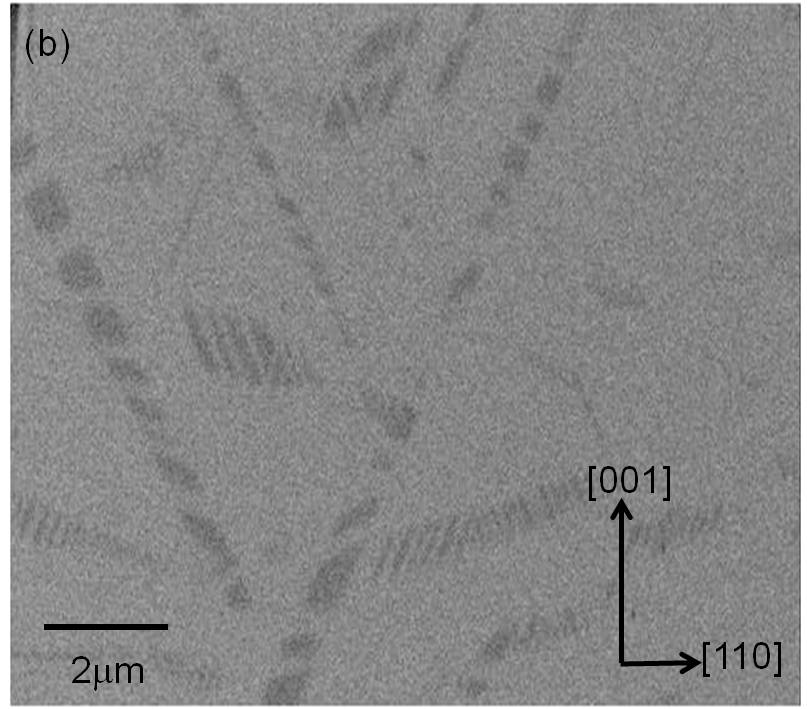}
  \end{center}
  \caption{Low-loss backscattered electron micrographs taken at 1kV on FIB cross-sections of crystal K100 on (a) \{100\} plane and (b) \{110\} plane.}
\label{fig:FIB}
\end{figure}

Figure \ref{fig:SE} shows typical secondary electron micrographs of freshly cleaved surfaces normal to the c-axis.  In addition to the terrace steps, square arrays of dark features with linear cross-section ($\textless 1\mu$m wide and 5-10$\mu$m long) can clearly be seen in all crystals.  EBSD orientation analysis indicates that these linear features are aligned along the crystallographic $\textless$110$\textgreater$ directions.  Crystal K100 has been cross-sectioned using a focussed ion beam microscope (FIB) along both (100) and (110) planes (see figure \ref{fig:FIB}).  This reveals that the minor phase consists of plates tilted at about 30$^{\circ}$ to [001], corresponding to \{113\} habit planes.  In addition, it can be seen from the micrographs in both plan view and cross-section that the plates are actually two-phase in nature, consisting of alternating platelets of dark and light contrast, each 100-200nm thick.  The habit planes of the platelets are also \{113\} in nature, with a plate aligned on the (113) plane consisting of platelets aligned on the (1$\bar{1}$3) plane. 

\subsection{Chemical analysis}
\begin{figure}[h!]
  \begin{center}
   \includegraphics[width=0.9\textwidth]{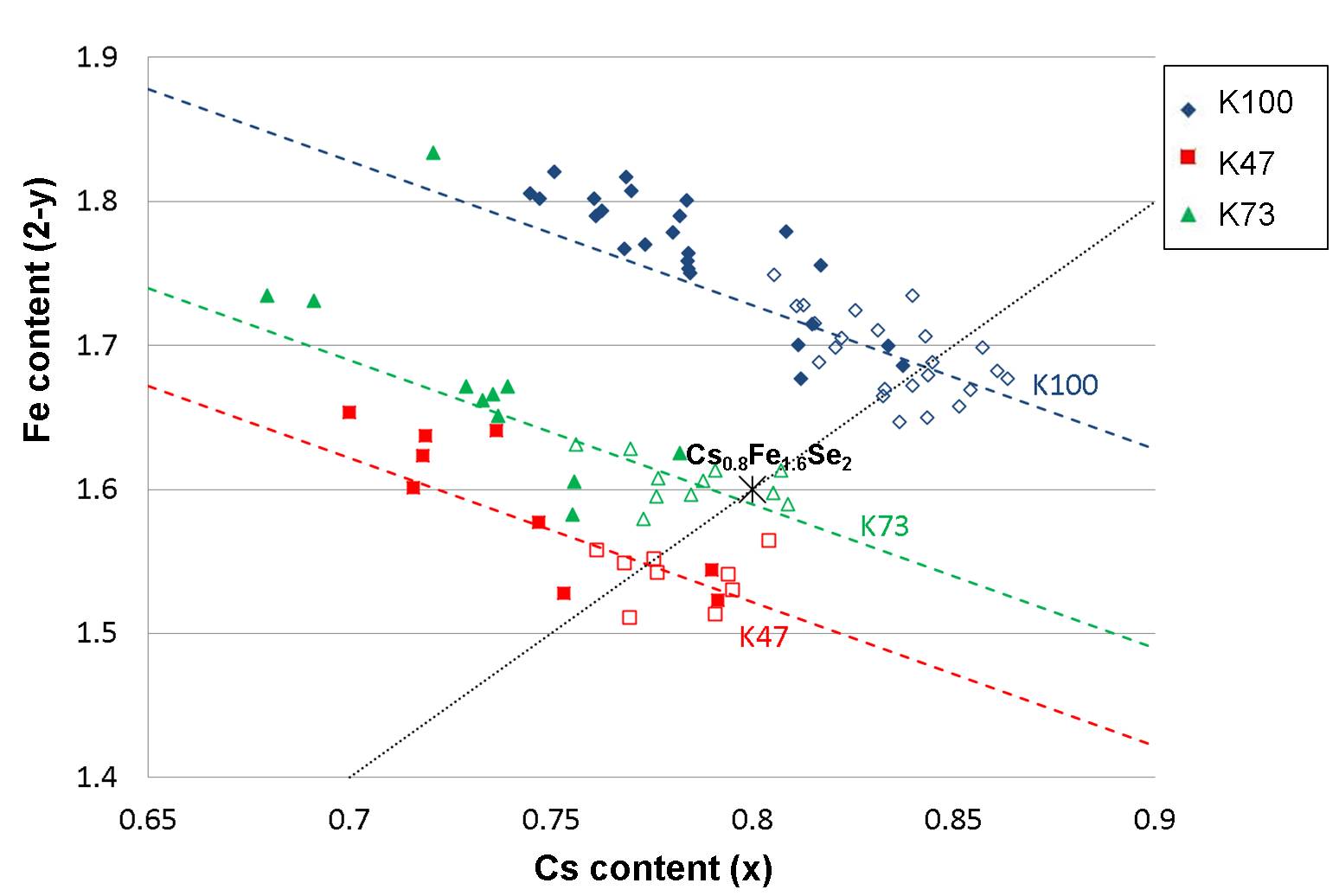}
  \end{center}
  \caption{Energy dispersive X-ray analysis of samples K100, K47 and K73.  The filled symbols are from the minority phase and the open symbols are from the matrix.}
\label{fig:EDX}
\end{figure}

Low-loss backscattered electron images taken at 1kV indicate that the minor phase is compositionally different to the matrix, with the minor phase appearing dark, corresponding to a composition with lower atomic weight.  Since Cs is significantly heavier than Fe or Se, the darker contrast suggests a decrease in Cs content in the minor phase compared to the matrix.  This has been confirmed with EDX analysis (given in figure \ref{fig:EDX}), which indicates that the minor phase in each crystal contains a larger Fe:Cs ratio than the matrix, although the quantitative data from the minor phase will underestimate the chemical variation as the features are smaller than the interaction volume from which the detected X-rays are generated.  Since XRD refinement suggests that there are no vacancies on the Se site \cite{Pomjakushin:2011}, the compositions in figure \ref{fig:EDX} have been expressed in terms of the Cs content (x) and the Fe content (2-y), assuming the Se content is 2.  For each of the three crystals, the compositions measured by EDX point analysis lie on straight lines with gradient $\approx$ -1, indicating that the different regions within a crystal have similar total cation content (i.e. $x+(2-y)=$constant) but with different Fe/Cs partitioning.   

The matrix compositions are also found to vary from sample to sample, with the Cs:Fe ratio remaining at about 1Cs:2Fe as shown by the dotted line in figure \ref{fig:EDX}.  K47 and K73 have matrix compositions very similar to the composition expected for the Fe vacancy ordered phase with $\sqrt5$x$\sqrt5$ superstructure, Cs$_{0.8}$Fe$_{1.6}$Se$_2$, but the matrix composition of the highest \Tc\ crystal, K100, is found  to be significantly different, with higher Cs and Fe content.  The matrix compositions of K47 and K73  measured by EDX are very similar to the average compositions found by micro-XRF (reported elsewhere \cite{Krzton-Maziopa:2012} and summarised in table \ref{table:samples}), but EDX analysis suggests that K73 contains slightly more Cs and Fe than K47, whereas the micro-XRF results found it to have a lower Fe content than K47.  There is a much larger discrepancy between the EDX and micro-XRF measurements for sample K100, with EDX analysis indicating that the matrix has considerably higher concentration of Fe and Cs than indicated by the average micro-XRF results.  
The origin of this discrepancy is not clear; whilst absolute concentrations measured by EDX must be treated with caution, the results presented here include data from at least two different flakes cleaved from each crystal, suggesting the differences in composition between the crystals can be interpreted with some confidence.  Macroscopic chemical variations within the crystal may be responsible, but all the  studies were carried out on samples taken from the region of each  crystal which appears  homogeneous in  micro-XRF maps \cite{Krzton-Maziopa:2012}. 


\subsection{HR-EBSD analysis}
\begin{figure}[h!]
  \begin{center}
   \includegraphics[width=0.7\textwidth]{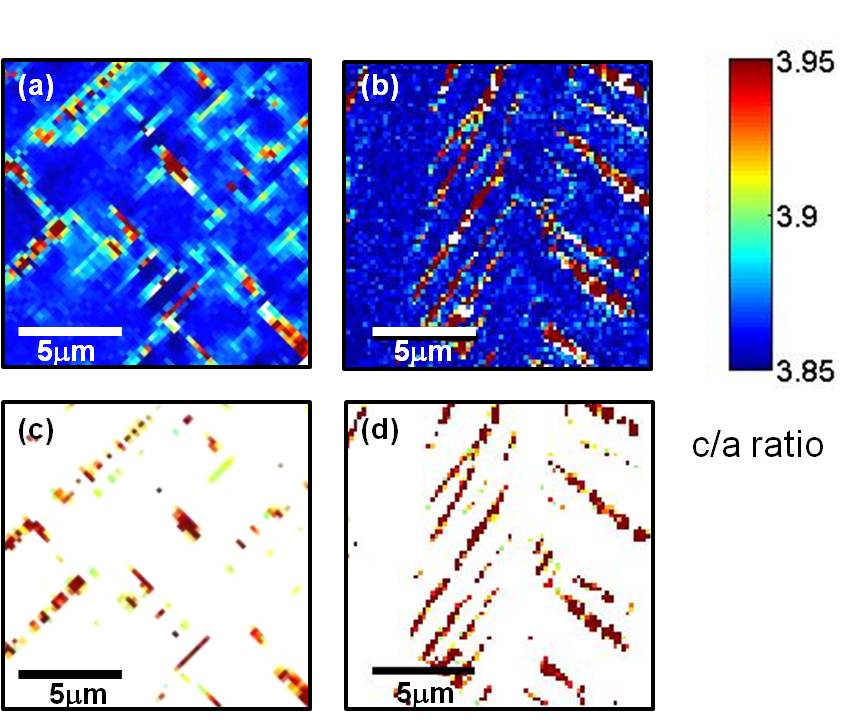}
  \end{center}
  \caption{HR-EBSD analysis showing variation in c/a ratio for samles (a) K100 and (b) K47.  Figures (c) and (d) show regions of the map with c/a$\textgreater$3.9 for samples K100 and K47 respectively.}
\label{fig:EBSD}
\end{figure}

Since accurate chemical analysis of the minor phase is difficult due to both the sensitivity of the EDX technique and the spatial resolution in bulk samples in the SEM, the HR-EBSD technique has been used to investigate small differences in the unit cell anisotropy (c/a ratio) between the matrix and the secondary phase.  Figure \ref{fig:EBSD}(a)and (b) show  HR-EBSD maps from typical regions of sample K100 and K47 respectively, with the colour coding indicating the local c/a ratio.  The absolute values of c/a have been obtained by setting the modal c/a value for the map to the bulk value obtained by powder X-ray diffraction  given in table \ref{table:samples}. In both crystals the minor phase is found to have substantially larger c/a ratio than the matrix, as clearly shown in figures \ref{fig:EBSD}(c) and (d) where only the pixels where c/a is greater than a threshold value of 3.9 are shown.  Similar features have been recently observed by Ricci \etal using scanning nano-focussed x-ray diffraction \cite{Ricci:2011b}, but the higher resolution EBSD results presented here reveals significantly more detail.  The areal fractions of the minor phase are found to be 6\%\ and 10\%\ in K100 and K47 respectively.   HR-EBSD mapping could not be carried out on crystal K73 because its insulating nature resulted in electrical charging in the scannning electron microscope and the surface sensitivity of this technique prevents high quality data being collected from carbon-coated specimens.  

The increase in c/a ratio found in the minor phase is consistent with the recent XRD study by Bosak \etal on the same crystal \cite{Bosak:2012}.  However, in the X-ray study the c-axis was found to expand by 2.4\% and the a-axis was found to contract by 2.4\%, resulting in an increase in c/a ratio of about 5\%.  In our study the average c/a values for the minor phases (i.e. regions shown in figures \ref{fig:EBSD}(c) and (d) where c/a ratio is larger than the threshold value of 3.9) correspond to an increase of only 1.7\%\ and 2.6\%\ in samples K100 and K47 respectively.  One explanation for this could be the presence of further nano-scale phase separation within the minority phase,  such as observed in high-resolution TEM studies on superconducting K$_x$Fe$_{2-y}$Se$_2$ crystals.  This would be consistent with the 3D reciprocal lattice XRD study in which
rod-like scattering parallel to (00L) associated with the minor phase is found, presumably originating from thin plate-shaped inclusions parallel to the a-b plane. Further studies using TEM are necessary to investigate this nanoscale phase separation.

\section{Influence of microstructure on superconducting properties}

The three crystals studied here all display antiferromagnetic ordering up to about 500K.  In addition, samples K100 and K47 also exhibit bulk superconductivity in magnetization measurements.  However, there is no clear trend between transition temperature and matrix composition in these crystals.  The matrix compositions of K47 and K73 are both very similar to the composition of the $\sqrt5$x$\sqrt5$ vacancy ordered Cs$_{0.8}$Fe$_{1.6}$Se$_{2}$ phase, despite K47 having a relatively high \Tc\ value of 23K whereas K73 is non-superconducting.  In contrast, sample K100, with the highest \Tc\ value of 27K, has a matrix composition richer in both Cs and Fe, whilst retaining the same 1:2 ratio of Cs to Fe.   The lack of a clear correlation between matrix composition and superconducting properties in our crystals suggest that the secondary phase rather than the matrix is superconducting.  Whilst the crystals contain only about 10$\%$ of the minor phase, our microstructural studies show that its morphology is such that a macroscopic percolation path could exist through the  minor phase, which would give a signature of  bulk superconductivity in magnetisation measurements.  This is consistent with very recent NMR results that indicate that the Fe vacancy free minor phase is superconducting and the ordered Fe vacancy major phase is an antiferromagnetic insulator \cite{Texier:2012}.  

It is interesting to note that the samples which exhibit superconductivity have higher starting Fe composition than the non-superconducting sample, and that the starting composition does not seem to strongly influence the composition of the major phase; for example K47 has significantly more Cs and Fe in the starting mixture than K73, but the actual matrix compositions are very similar in both crystals.  It is possible that the excess Fe in K47 makes the minority phase richer in Fe in K47 compared with K73, producing  the difference in superconducting properties, but the EDX analysis is not able to give sufficiently accurate compositional information on the minority phase to confirm this.  

HR-EBSD analysis has been carried out in addition to the chemical analysis to investigate the structural variations between the two phases.  The minority phase in samples K100 and K47 crystals is found to have a higher c/a ratio compared to the matrix.  Whilst sample K47 has a smaller bulk c/a ratio than K100 (corresponding to lower Cs and Fe content), the increase in c/a ratio of the minor phase compared to the matrix is greater in sample K47 than in K100, resulting in similar average c/a ratio values for the minority phases of the superconducting samples (3.96 and 3.93 respectively). This supports the hypothesis that the minor phase is responsible for superconductivity, since the transition temperatures for K100 and K47 are similar.  



\section{Comparison with Fe(Se,Te) system}

\begin{figure}[h!]
  \begin{center}
   \includegraphics[width=0.9\textwidth]{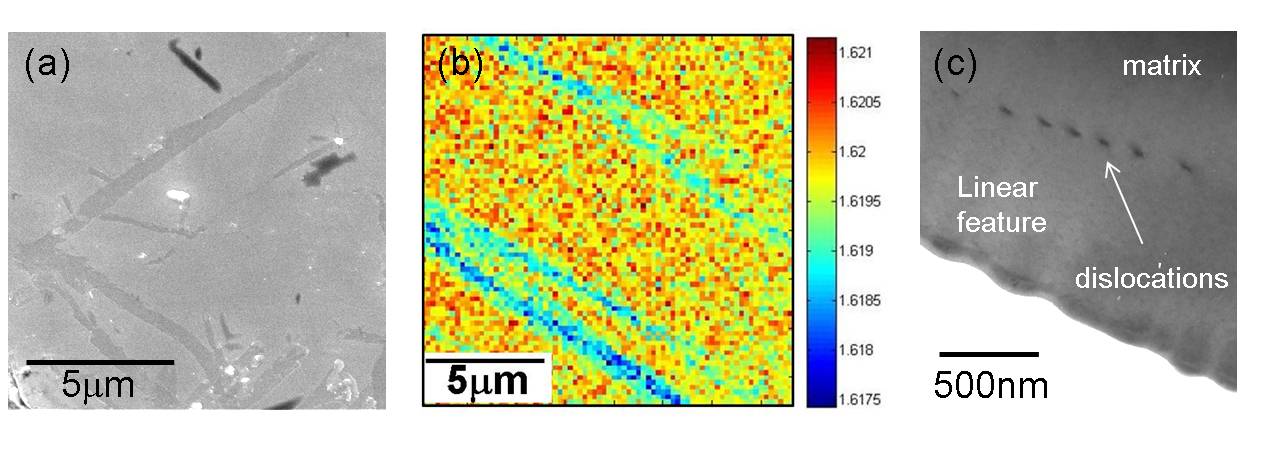}
  \end{center}
  \caption{Microsctructural analysis of Fe$_{1.07}$Se$_{0.25}$Te$_{0.75}$ single crystal showing (a) secondary electron micrograph, (b) HR-EBSD map showing variation in c/a ratio and (c) TEM image of a cross-section through one of the linear features prepared by a focussed ion beam microscope.}
\label{fig:FST}
\end{figure}

Fe(Se,Te) compounds also exhibit magnetism co-existing with superconductivity over certain ranges of composition.  Microstructural studies on a range of  Fe$_y$Se$_{0.25}$Te$_{0.75}$ compounds containing varying amounts of iron have shown that small spatial variations in chemical composition, with related lattice parameter variations, are present in the samples which exhibit both superconductivity and magnetism, whereas the more homogeneous crystals were purely magnetic in nature \cite{Speller:2011}.  In contrast to the \CFS\ crystals, the microstructure of the two-phase samples is rather different, consisting of smaller composition and structural variations in much larger domains (25-50 microns).  However smaller, linear-shaped features were also found in these crystals, with a particularly high density present in the crystal grown from the most Fe-rich starting composition (as shown in figure \ref{fig:FST}(a)).  HR-EBSD analysis on these features show a small decrease in c/a ratio, with TEM/EDX analysis confirming that they are slightly richer in Fe than the matrix.  The TEM micrograph shown in figure \ref{fig:FST} (b) shows that the interface between the  matrix and the linear features consists of a row of dislocations, presumably accommodating the lattice mismatch and/or slight angular misorientation.  These features differ substantially from those found in the \CFS\ samples since they are rod-like, single-phase features, only slightly different in composition and lattice parameter to the matrix.  In contrast the \CFS\ samples presented here contain two-phase, plate-like features in which one of the phases is significantly different structurally and chemically to the matrix.  The features found in the Fe(Se,Te) samples are most prevalent in the insulating samples, whereas the minority phase in our \CFS\ crystals are thought to be superconducting.  

In both ternary compounds, the local variations in unit cell anisotropy (c/a ratio) on the microscopic scale revealed by HR-EBSD mapping are believed to influence the  superconducting properties.  The volume fraction of superconducting phase in Fe(Se,Te) crystals is found to be associated with the fraction of the crystal with c/a below a certain value, corresponding to structures containing none of the excess Fe which is found to be detrimental to superconductivity.  In the case of \CFS\ samples, superconductivity is suppressed by the presence of Fe vacancies characterised by a decrease in the c/a ratio.  Since sp vacancy ordered phases with a well-defined vacancy concentration are stable in the \CFS\ compounds, clear phase separation is found in these compounds.  In contrast, Fe(Se,Te) compounds can support a continuous range of excess Fe concentrations within the same phase.

\section{Summary}
In order to understand the interplay between superconductivity and magnetism in Fe-based compounds, it is essential to investigate the structure of the crystals on a range of different length scales.  TEM and STM studies have recently revealed the interesting multiphasic nature of these crystals on the nano-scale, and here we have presented complementary studies on the micron  scale, clearly demonstrating that \CFS\ crystals have complex microstructures, consisting of a network of crystallographically-aligned plates containing two different phases distributed throughout the matrix.  The chemical and structural analysis is consistent with the minor Fe-rich being responsible for superconductivity in these crystals.  The phase separation in the \CFS\ samples is more dramatic than the chemical and structural inhomogeneities found in Fe(Se,Te) samples, but in both cases chemical inhomogeneities are thought to be responsible for the coexistence of magnetism with superconductivity. 
 
\ack{Dr. S Speller is supported under the Royal Academy of Engineering / EPSRC Fellowship scheme.}

\section*{References}
\bibliographystyle{unsrt}
\bibliography{CFS}

\end{document}